\pdfoutput=1

\documentclass[twocolumn,showpacs]{revtex4}
\usepackage{amsmath,amssymb,graphicx}
\begin{document}

\title{
  Quantum-limited shot noise and quantum interference 
  in graphene based Corbino disk
}

\author{Grzegorz Rut}
\affiliation{Marian Smoluchowski Institute of Physics, 
Jagiellonian University, Reymonta 4, PL--30059 Krak\'{o}w, Poland}
\author{Adam Rycerz}
\affiliation{Marian Smoluchowski Institute of Physics, 
Jagiellonian University, Reymonta 4, PL--30059 Krak\'{o}w, Poland}

\begin{abstract}
This is a theoretical study of finite voltage effects on the conductance, the shot noise power, and the third charge-transfer cumulant for graphene-based Corbino disk in the presence of external magnetic fields. Periodic magnetoconductance oscillations, predicted in Refs.\ \cite{Ryc10,Kat10}, become invisible for relatively small source-drain voltages, as the current decays rapidly with magnetic field. Quantum interference still governs the behavior of higher charge-transfer cumulants. 
\end{abstract}

\date{\today}
\maketitle

\section{Introduction}
\begin{figure}[!t]
  \centerline{\includegraphics[width=0.9\linewidth]{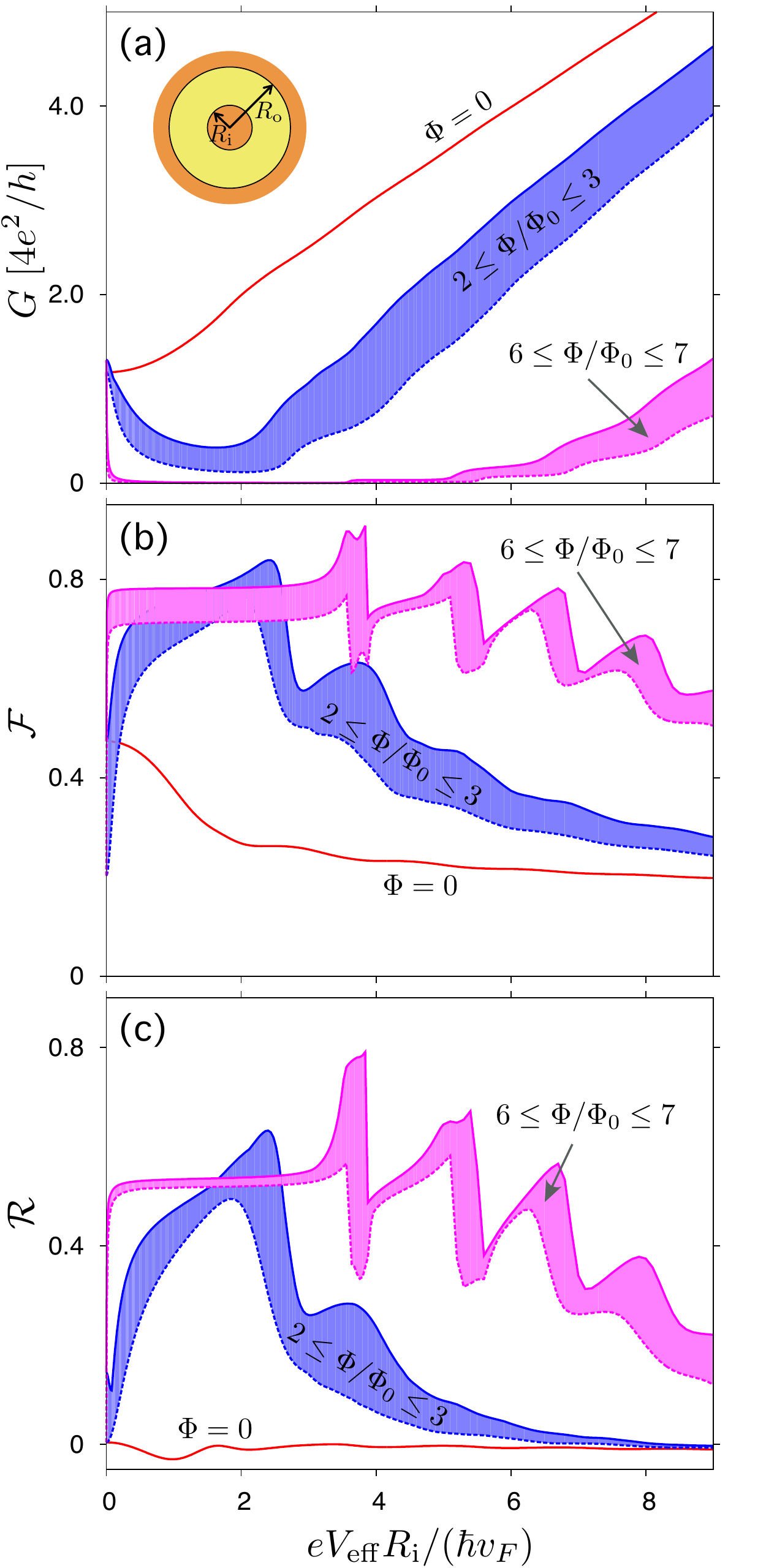}}
  \caption{\label{figvsvol}
    Variation ranges for the finite-voltage conductance (a), Fano factor (b), and ${\cal R}$-factor (c) in cases the magnetic flux $\Phi$ piercing the disk area $R_{\rm i}<r<R_{\rm o}$ [see inset in panel (a)] is varied in the limits given by Eq.\ (\ref{phimphi}) with $m_\Phi=3$ and $m_\Phi=7$. The values for $\Phi=0$ are also shown.
  }
\end{figure}

The Corbino geometry, in which electric current is passed through a~disk-shaped sample area attached to two circular leads [see Fig.~\ref{figvsvol}(a), the inset], was proposed over a~century ago \cite{Bol86,Ada15} to measure the magnetoresistance without generating the Hall voltage, making a~significant step towards understanding the nature of charge transport in ordinary solids \cite{Gal91}. An interest in such a~geometry has reappeared due to the fabrication of GaAs/AlGaAs heterostructures \cite{Kir94,Man96}.
Also, after the discovery of high-temperature superconductivity, Corbino measurements have provided a~valuable insight into the vortex dynamics, as the influence of sample edges was eliminated \cite{Ryc99}. 

In the context of graphene, various transport properties of Corbino disks were recently studied experimentally \cite{Yan10,Fau10,Zha12} and theoretically \cite{Ryc10,Kat10,Kha13}. Next to the case of a~rectangular strip geometry \cite{Two06,Pra07}, the Corbino geometry provides another situation when transport properties of a~graphene nanodevice can be investigated analytically \cite{Ryc10}, by solving the scattering problem for Dirac fermions, at arbitrary dopings and magnetic fields. In particular, mode-matching analysis for the effective Dirac equation gives transmission probabilities for {\em undoped} disk in graphene monolayer \cite{Ryc10,Kat10}
\begin{equation}
  \label{tjzero}
  T_j^{(0)}=\frac{1}{\cosh^2\left[(j+\Phi/\Phi_0)\ln(R_{\rm o}/R_{\rm i})\right]},
\end{equation}
where $j=\pm{}\frac{1}{2},\pm{}\frac{3}{2},\dots$ is the angular-momentum quantum number labeling normal modes in the leads, $\Phi=\pi(R_\mathrm{o}^2-R_\mathrm{i}^2)B$ is the flux piercing the disk in the uniform magnetic field ($B$), $R_{\rm i}$ is the inner radius , $R_{\rm o}$ is the outer radius, and $\Phi_0=2\,(h/e)\ln(R_{\rm o}/R_{\rm i})$. Moreover, in the derivation of Eq.\ (\ref{tjzero}),  the limit of heavily-doped graphene leads  \cite{Two06} is imposed.
Summing $T_j$-s over the normal modes in the leads, one finds that the Landauer-B\"{u}ttiker conductance, in the linear-response regime, shows periodic (approximately sinusoidal) oscillations with the flux $\Phi$, with $\Phi_0$ being the oscillations period. Additionally, the disk conductance averaged over a~single period restores the pseudodiffusive value \cite{Ryc09}
\begin{equation}
  G_{\rm diff}=\frac{2\pi\sigma_0}{\ln(R_{\rm o}/R_{\rm i})},
\end{equation}
with $\sigma_0=(4/\pi)\,e^2/h$ being the universal conductivity of graphene. Analogous behavior is predicted for higher charge-transfer cumulants \cite{Kat10,Ryc12}.

In this paper, we extend the analysis beyond the linear-response regime by calculating the conductance, the Fano factor ${\cal F}$ quantifying the shot-noise power, and ${\cal R}$-factor quantifying the third charge-transfer cumulant, in a~situation when finite source-drain voltage is applied to graphene-based Corbino disk in the shot-noise limit. Our results show that albeit the conductance oscillations vanish rapidly with the voltage and magnetic field, ${\cal F}$ and ${\cal R}$ still oscillate periodically and their mean values approach $\overline{\cal F}_\infty\simeq{}0.76$ and $\overline{\cal R}_\infty\simeq{}0.55$ (respectively) in the high-field limit. In the remaining parts of the paper, we first (in Section~II) recall briefly the formula allowing one to determine the conductance and other charge-transfer characteristics of graphene-based Corbino disk at arbitrary voltages and magnetic fields. Next, in Section~III, our numerical results are presented in details. The conclussions are given in Section~IV.

\section{Charge-transfer cumulants}
In the shot-noise limit $eV_{\rm eff}\gg{}k_BT$, with $V_{\rm eff}$ being the effective source-drain voltage \cite{voltfoo}, the charge $Q$ passing a~nanoscale device in a~time interval $\Delta{}t$ is a~random variable, with the characteristic function $\Lambda(\chi)$ given by the Levitov formula \cite{Naz09}
\begin{multline}
  \ln\Lambda(\chi)\equiv{}\ln\left<\exp\left(i\chi{}Q/e\right)\right>= \\
  4_{(\sigma,v)}\frac{\Delta{}t}{h}
  \int\limits_{\mu_0-\frac{eV_{\rm eff}}{2}}^{\mu_0+\frac{eV_{\rm eff}}{2}}
  d\epsilon\,\sum_j\ln\left[
    1+\left(e^{i\chi}\!-\!1\right)T_j(\epsilon)
  \right],
\end{multline}
where $\langle{}X\rangle$ denotes the expectation value of $X$, the factor $4_{(\sigma,v)}$ accounts for spin and valley degeneracies, and we have assumed $V_{\rm eff}>0$ without loss of a~generality. The average charge $\langle{Q}\rangle$, as well as any charge-transfer cumulant $\langle\langle{Q^m}\rangle\rangle\equiv\langle\,(Q-\langle{Q}\rangle)^m\,\rangle$, may be obtained by subsequent differentiation of $\ln\Lambda(\chi)$ with respect to $i\chi$ at $\chi=0$. In particular, the conductance 
\begin{multline}
  \label{gfinvo}
  G(V_{\rm eff})=\frac{\langle{Q}\rangle}{V_{\rm eff}\Delta{t}}=
  \frac{e}{V_{\rm eff}\Delta{t}}
  \left.\frac{\partial\ln\Lambda}{\partial(i\chi)}\right|_{\chi=0} \\
  \equiv \frac{4_{(\sigma,v)}e^2}{h}\,\sum_j
  \big<T_j\big>_{|\epsilon-\mu_0|\leqslant{}\frac{eV_{\rm eff}}{2}},
\end{multline}
where transmission probabilities $T_j(\epsilon)$ are averaged over the energy interval $|\epsilon-\mu_0|\leqslant{}eV_{\rm eff}/2$. Analogously,
\begin{multline}
  \label{ffinvo}
  {\cal F}(V_{\rm eff})={\displaystyle\langle\langle{Q^2}\rangle\rangle}\big/{\displaystyle\langle\langle{Q^2}\rangle\rangle_{\rm Poisson}} \\
=\frac{\sum_j\big<T_j\left(1-T_j\right)\big>_{|\epsilon-\mu_0|\leqslant{}\frac{eV_{\rm eff}}{2}}}{\sum_j\big<T_j\big>_{|\epsilon-\mu_0|\leqslant{}\frac{eV_{\rm eff}}{2}}}
\end{multline}
and
\begin{multline}
  \label{rfinvo}
  {\cal R}(V_{\rm eff})=\langle\langle{Q^3}\rangle\rangle\big/\langle\langle{Q^3}\rangle\rangle_{\rm Poisson} \\
=\frac{\sum_j\big<T\left(1-T_j\right)\left(1-2T_j\right)\big>_{|\epsilon-\mu_0|\leqslant{}\frac{eV_{\rm eff}}{2}}}{\sum_j\big<T_j\big>_{|\epsilon-\mu_0|\leqslant{}\frac{eV_{\rm eff}}{2}}},
\end{multline}
with $\langle\langle{Q^m}\rangle\rangle_{\rm Poisson}$ the value of $m$-th cumulant for the Poissonian limit ($\,T_j(\epsilon)\ll{}1\,$), given by a~generalized Schottky formula 
$\langle\langle{Q^m}\rangle\rangle_{\rm Poisson}= e^{m-1}\langle{Q}\rangle$. 

In the case of graphene-based Corbino disk, the energy-dependent transmission probabilities are given by \cite{Ryc10}
\begin{equation}
\label{tjdop}
  T_j(\epsilon)=
  \frac{16\,(\tilde{\epsilon}^2/\beta)^{|2j-1|}}%
{\tilde{\epsilon}^2R_\mathrm{i}R_\mathrm{o}\,(X_j^2+Y_j^2)}
  \left[\frac{\Gamma(\gamma_{j\uparrow})}{\Gamma(\alpha_{j\uparrow})}\right]^2,
\end{equation}
where $\tilde{\epsilon}=\epsilon/(\hbar{}v_F)$ with $v_F\simeq{}10^6\,$m/s the energy-independent Fermi velocity, $\beta=eB/(2\hbar)$, $\Gamma(z)$ is the Euler Gamma function, and 
\begin{align}
  \alpha_{js} &= \frac{1}{4}\left[\,2(j+m_s+|j-m_s|+1)
  -\frac{\tilde{\epsilon}^2}{\beta}\,\right], 
  \nonumber\\
  \gamma_{js} &= |j-m_s|+1,
\end{align}
with  $m_s=\pm\frac{1}{2}$ for the lattice pseudospin $s=\uparrow,\downarrow$. The remaining symbols in Eq.\ (\ref{tjdop}) are defined as
\begin{align}
  X_j &= w_{j\uparrow\uparrow}^- + z_{j,1}z_{j,2}w_{j\downarrow\downarrow}^-, \nonumber\\
  Y_j &= z_{j,2}w_{j\uparrow\downarrow}^+ - z_{j,1}w_{j\downarrow\uparrow}^+, 
\end{align}
where
\begin{align}
  w_{jss'}^\pm &= \xi_{js}^{(1)}(R_\mathrm{i})\xi_{js'}^{(2)}(R_\mathrm{o})
  \pm \xi_{js}^{(1)}(R_\mathrm{o})\xi_{js'}^{(2)}(R_\mathrm{i}), \nonumber\\
  z_{j,1} &= [2(j+s_j)]^{-2s_j}, \label{wzzdefs}\\
  z_{j,2} &= 2(\beta/\tilde{\epsilon}^2)^{s_j+1/2}, \nonumber
\end{align}
with $s_j\equiv\frac{1}{2}\mbox{sgn}(j)$. The functions $\xi_{js}^{(1)}(r)$ and $\xi_{js}^{(2)}(r)$ in the first line of Eq.\ (\ref{wzzdefs}) are given by
\begin{multline}
  \label{xisnu}
  \xi_{js}^{(\nu)}(r)=(\tilde{\epsilon}r)^{|j-m_s|}\exp({-\beta{r}^2/2}) \\
  \times
  \left\{\begin{array}{cc}
      M(\alpha_{js},\gamma_{js},\beta{r}^2), & \text{if}\ \ \nu\!=\!1, \\ 
      U(\alpha_{js},\gamma_{js},\beta{r}^2), & \text{it}\ \ \nu\!=\!2, \\
    \end{array}\right.
\end{multline}
where $M(a,b,z)$ and $U(a,b,z)$ are the confluent hypergeometric functions \cite{Abram,bsignfoo}. 

It can be shown that in the zero-energy limit $(\epsilon\rightarrow{}0)$ Eq.\ (\ref{tjdop}) simplifies to Eq.\ (\ref{tjzero}). Similarly, in case the energy is adjusted to a~higher Landau level (LL), namely, $\tilde{\epsilon}^2/(4\beta)=n=1,2,\dots$, the transmission probability for $j$-th normal mode (in the high-field limit) is $T_j^{(n)}=T_{j-2n}^{(0)}$ \cite{Ryc10}. In effect, periodic magnetoconductance oscillations in the linear-response regime are followed by similar oscillations of ${\cal F}$ and ${\cal R}$ (for analytic Fourier decompositions, see Ref.\ \cite{Ryc12}), with the mean values
\begin{equation}
  \label{frpdiff}
  {\cal F}_{\rm diff}=1/3\ \ \ \ \text{and}\ \ \ \ {\cal R}_{\rm diff}=1/15,
\end{equation}
provided the disk is undoped or the doping is adjusted to any higher LL. These are the basic features of a~nonstandard quantum interference phenomenon, which may appear when charge transport in graphene (or other Dirac system) is primarily carried by evanescent modes \cite{Kol12}.

\begin{figure}[!t]
  \centerline{\includegraphics[width=0.9\linewidth]{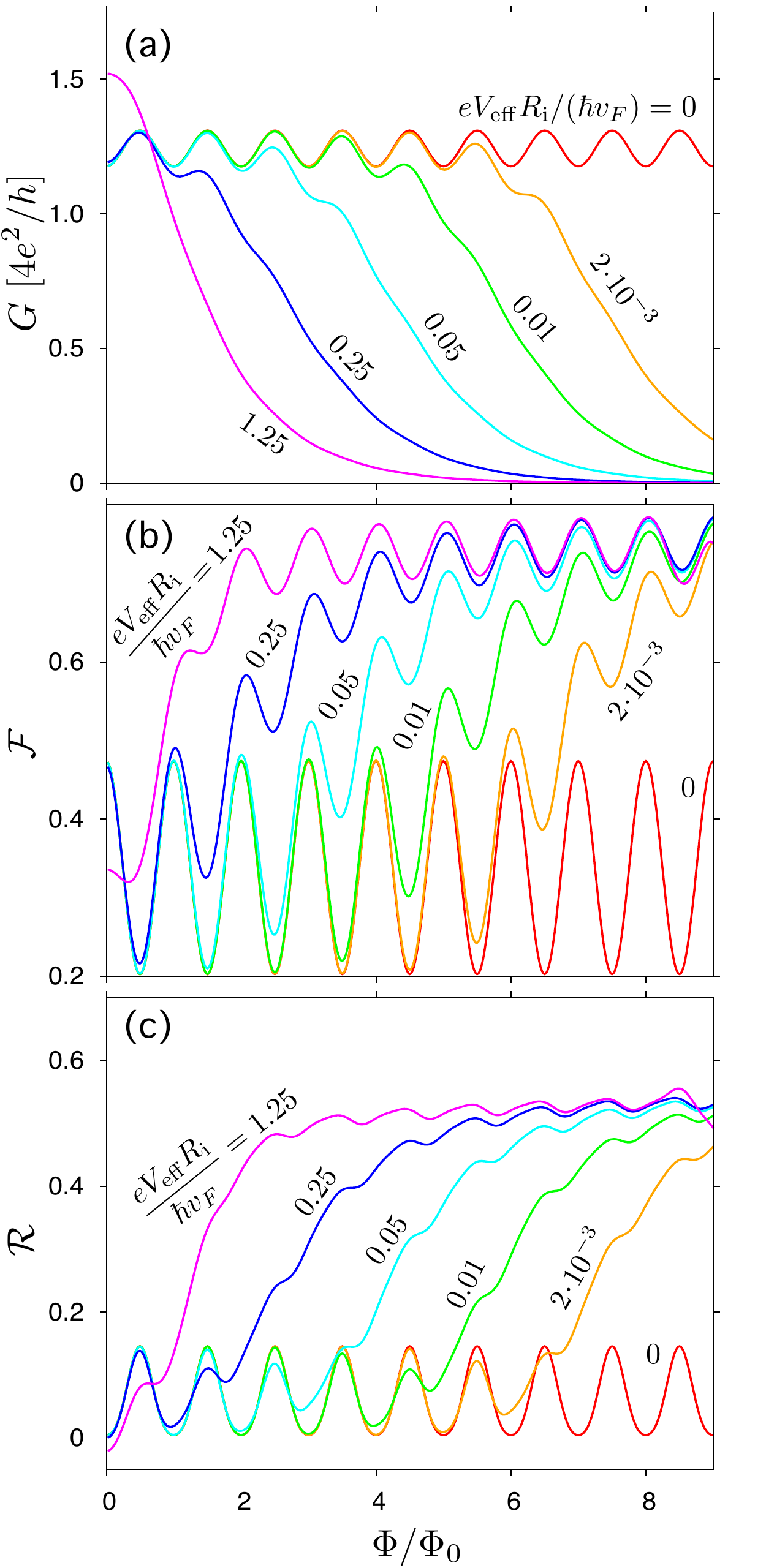}}
  \caption{\label{figvsphi}
    Magnetic flux effect on the finite-voltage conductance (a), Fano factor (b), and ${\cal R}$-factor (c). The effective source-drain voltage $V_{\rm eff}$ is specified for each curve.
  }
\end{figure}

\begin{figure}[!t]
  \centerline{\includegraphics[width=0.9\linewidth]{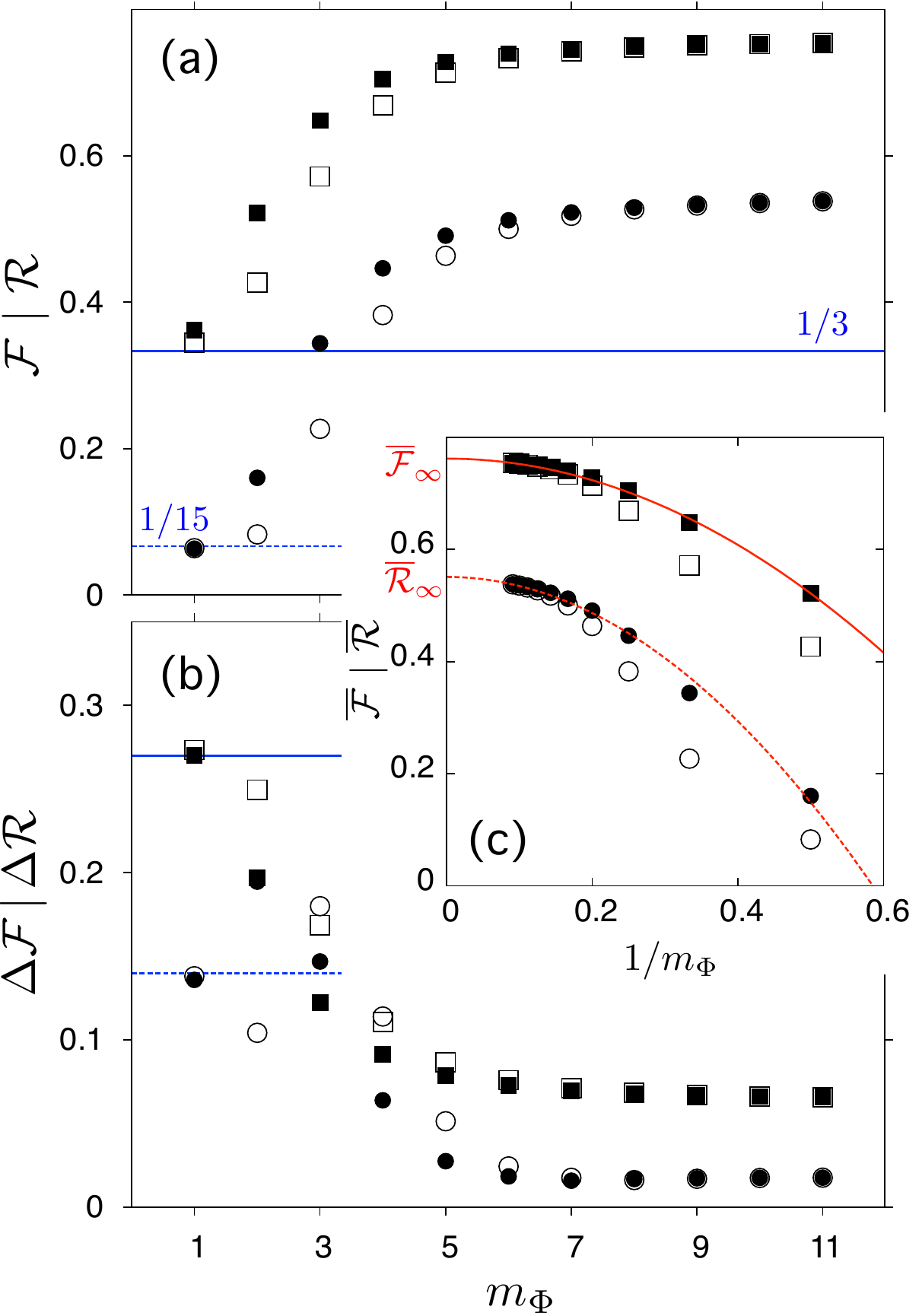}}
  \caption{\label{figsteps}
    Average values $\overline{X}$ (a) and oscillation magnitudes $\Delta{X}={\rm max}\,(X)-{\rm min}\,(X)$ (b), with $X={\cal F}$ (squares) and $X={\cal R}$ (circles), calculated for several consecutive flux intervals defined by Eq.\ (\ref{phimphi}). Open (or closed) symbols at each panel correspond to $eV_{\rm eff}R_{\rm i}/(\hbar{}v_F)=0.25$ (or $0.5$). Lines in panels (a) and (b) depict the linear-response values given by Eqs.\ (\ref{frpdiff}) and (\ref{delxlin}). Panel (c) illustrates the scaling of $\overline{\cal F}$ and $\overline{\cal R}$ with $1/m_\Phi\rightarrow{}0$ (see the main text for details). 
  }
\end{figure}

\section{Results and discussion}
Several factors, not taken into account in the above analysis, may make it difficult to confirm experimentally the effects which are described in Refs.\ \cite{Ryc10,Kat10}. These include the influence of disorder, electron-phonon coupling, or electron-electron interactions; i.e., the factors which are absent (or suppressed) in several analogs of graphene \cite{Sin11,Liu14,Bor14}, and which are beyond the scope of this paper. Another potential obstacle is related to the fact that resonances with distinct LLs shrink rapidly with increasing field, making the linear-response regime hard to access. Here we point out that theoretical discussion of charge transport through graphene-based Corbino disk still can be carried out, in a~rigorous manner, beyond the linear-response regime.

For the purpose of numerical demonstration, we choose $R_{\rm o}/R_{\rm i}=5$, and focus on the~vicinity of the Dirac point by setting $\mu_0=0$ \cite{mu0foo}. The corresponding oscillation magnitudes, in the linear-response limit, are \cite{Ryc12}
\begin{gather}
  \label{delxlin}
  \Delta{}G(V_{\rm eff}\rightarrow{}0)=0.11\,G_{\rm diff}, \\
  \Delta{\cal F}(V_{\rm eff}\rightarrow{}0)=0.27, 
  \ \ \ \ 
  \Delta{\cal R}(V_{\rm eff}\rightarrow{}0)=0.14. \nonumber
\end{gather}
For any finite $V_{\rm eff}$ and any flux $\Phi$, the averages in Eqs.\ (\ref{gfinvo}), (\ref{ffinvo}), and (\ref{rfinvo}) can be calculated numerically after substituting $T_j(\epsilon)$ given by Eq.\ (\ref{tjdop}). Our main results are presented in Figs.~\ref{figvsvol}, \ref{figvsphi}, and \ref{figsteps}. 

First, in Figs.~\ref{figvsvol}(a)--(c), we have depicted the values taken by $G(V_{\rm eff})$, ${\cal F}(V_{\rm eff})$, and ${\cal R}(V_{\rm eff})$, when the flux is varied in separate intervals, each of which having $\Phi_0$ width, namely
\begin{equation}
  \label{phimphi}
  (m_\Phi\!-\!1)\,\Phi_0\leq{}\Phi\leq{}m_\Phi\Phi_0, \ \ \ \ m_\Phi=1,2,\dots.
\end{equation}
The two shaded areas are for $m_\Phi=3$ and $m_\Phi=7$; distinct solid line (at each panel) depicts the corresponding charge-transfer characteristic at $\Phi=0$. It is clear from Fig.~\ref{figvsvol} that $G(V_{\rm eff})$ is strongly suppresed by the magnetic field provided that $eV_{\rm eff}\lesssim{}\hbar{}v_F/R_{\rm i}$. For higher $V_{\rm eff}$ the ballistic transport regime is entered, leading to $G(V_{\rm eff})\propto{}V_{\rm eff}$, ${\cal F}(V_{\rm eff})\lesssim{}0.2$, and ${\cal R}(V_{\rm eff})\simeq{}0$ in the $eV_{\rm eff}\gg{}\hbar{}v_F/R_{\rm i}$ limit. Most remarkably, for $0<eV_{\rm eff}\lesssim{}\hbar{}v_F/R_{\rm i}$ and the highest discussed flux interval ($m_\Phi=7$),  ${\cal F}(V_{\rm eff})$ and  ${\cal R}(V_{\rm eff})$ take the values from narrow ranges around ${\cal F}\simeq{0.7}$ and ${\cal R}\simeq{}0.5$ [see Figs.~\ref{figvsvol}(b) and \ref{figvsvol}(c)], coinciding with recent findings for transport near LLs in graphene bilayer \cite{Rut14}.

\begin{table}
  \caption{ \label{tabinfty} 
    Limiting values of period-averaged $\overline{\cal F}$, $\overline{\cal R}$ and oscillation magnitudes $\Delta{}\cal F$, $\Delta{}\cal R$ obtained by least-squares fitting of the parameters in Eq.\ (\ref{ymphi}). Numbers in parentheses are standard deviations for the last digit (see also Ref.\ \cite{fooinfty}). 
  }
  \begin{tabular}{c|cc|cc}
    \hline\hline
    $R_{\rm o}/R_{\rm i}$
    & $\overline{\cal F}_\infty$ & $\Delta{\cal F}_\infty$
    & $\overline{\cal R}_\infty$ & $\Delta{\cal R}_\infty$  \\ \hline
    $2.5$ & $\ $0.761(1) & $\,$0.0014(1) & $\ $0.552(3) & $\,$0.0064(2) \\
    $5.0$ & $\ $0.763(1) & $\,$0.061(1)$\,\ $ & $\ $0.555(2)
    & $\,$0.017(1)$\,\ $ \\
    $10$ & $\ $0.771(5) & $\,$0.191(2)$\,\ $ & $\ $0.56(1)$\,\ $
    & $\,$0.170(2)$\,\ $ \\
    \hline\hline
  \end{tabular}
\end{table}

These observations are further supported with the data presented in Fig.~\ref{figvsphi}, where the conductance and other charge-transfer characteristics are plotted directly as functions of $\Phi$, for selected values of $V_{\rm eff}$. Although $G(V_{\rm eff})$ decays relatively fast with $\Phi$ for any $V_{\rm eff}\neq{}0$, such that magnetoconductance oscillations are visible for $eV_{\rm eff}\ll{}\hbar{}v_F/R_{\rm i}$ only [see Fig.\ \ref{figvsphi}(a)], ${\cal F}(V_{\rm eff})$ and ${\cal R}(V_{\rm eff})$ show periodic oscillations at high fields for arbitrary $V_{\rm eff}$ [see Figs.\ \ref{figvsphi}(b) and \ref{figvsphi}(c)]. In order to describe these oscillations in a~quantitative manner, we have calculated numerically the average values of ${\cal F}(V_{\rm eff})$ and ${\cal R}(V_{\rm eff})$, as well as the corresponding oscillation magnitudes, for several consecutive flux intervals defined by Eq.\ (\ref{phimphi}), and depicted them as functions of the interval number ($m_\Phi$) in Figs.\ \ref{figsteps}(a), \ref{figsteps}(b). Next, the scaling with $1/m_\Phi\rightarrow{}0$ is performed by least-squares fitting of the approximating formula
\begin{equation}
  \label{ymphi}
  Y\left[m_\Phi\right] \simeq Y_\infty+A_Y\left(\frac{1}{m_\Phi}\right)^2,
\end{equation}
for $Y=\overline{\cal F}$, $\overline{\cal R}$, $\Delta{\cal F}$, and $\Delta{\cal R}$. The examples of $\overline{\cal F}[m_\Phi]$ and  $\overline{\cal R}[m_\Phi]$ are presented in Fig.\ \ref{figsteps}(a); the values of $Y_\infty$ for different ratios $R_{\rm o}/R_{\rm i}$ are listed in Table~\ref{tabinfty}  \cite{fooinfty}. 

A striking feature of the results presented in Table~\ref{tabinfty} is the total lack of effects of both the radii ratio $R_{\rm o}/R_{\rm i}$ and the source-drain voltage $V_{\rm eff}$ on limiting values of $\overline{\cal F}_\infty$ and  $\overline{\cal R}_\infty$. (In contrast, $\Delta{\cal F}_\infty$ and $\Delta{\cal R}_\infty$ strongly depends on $R_{\rm o}/R_{\rm i}$.) This fact allows us to expect the quantum-limited shot noise, characterized by 
\begin{equation}
  \overline{\cal F}_\infty\simeq{}0.76\ \ \ 
  \text{and}\ \ \ 
  \overline{\cal R}_\infty\simeq{}0.55,
\end{equation}
to appear generically in graphene-based nanosystems at high magnetic fields and for finite source-drain voltages, similarly as pseudodiffusive shot-noise (with ${\cal F}_{\rm diff}=1/3$ and  ${\cal R}_{\rm diff}=1/15$) appears generically at the Dirac point in the linear-response limit.

\section{Conclusions}
We have investigated the finite-voltage effects on the magnetoconductance, as well as the magnetic-field dependence of the shot-noise power and the third charge-transfer cumulant, for the Corbino disk in ballistic graphene. Periodic magnetoconductance oscillations, earlier discussed theoretically in the linear-response limit \cite{Ryc10,Kat10}, are found to decay rapidly with increasing field at finite voltages. To the contrary, the ${\cal F}$ and ${\cal R}$-factors, quantifying the higher charge-transfer cumulants, show periodic oscillations for arbitrary high fields, for both the linear-response limit and the finite-voltage case. Although such oscillations must be regarded as signatures of a~nonstandard quantum interference phenomena, specific for graphene-based disks near zero doping (and having counterparts for higher Landau levels), the parameter-independent mean values of  $\overline{\cal F}_\infty\simeq{}0.76$ and $\overline{\cal R}_\infty\simeq{}0.55$ suggest the existence of a~generic, finite-voltage and high-field analog of a~familiar pseudodiffusive charge transport regime in ballistic graphene. 

We hope our findings will motivate some experimental attempts to understand the peculiar nature of quantum transport via evanescent waves in graphene, which manifests itself not only in the well-elaborated multimode case of wide rectangular samples \cite{Two06,Pra07,Ryc09}, but also when a~very limited number of normal modes contribute to the system conductance and other charge-transfer characteristics, as in the case of Corbino disks with large radii ratios $R_{\rm o}/R_{\rm i}\gg{}1$. Albeit the discussion is, in principle, limited to the system with a perfect circular symmetry and the uniform magnetic field, special features of the results, in particular the fact that mean values of the ${\cal F}$ and ${\cal R}$-factors are insensitive to the radii ratio and to the voltage, allow us to believe that quantum-limited shot noise as well as the signatures of quantum interference should appear in more general situations as well.

\section*{Acknowledgments}
The work was supported by the National Science Centre of Poland (NCN)
via Grant No.\ N--N202--031440, and partly by Foundation for Polish Science
(FNP) under the program TEAM. The computations were partly performed using 
the PL-Grid infrastructure.


\begin{thebibliography}{}

\bibitem{Ryc10}
A.~Rycerz, Phys.\ Rev.\ B {\bf 81}, 121404(R) (2010).

\bibitem{Kat10}
M.I.~Katsnelson, Europhys.\ Lett.\ \textbf{89}, 17001 (2010).

\bibitem{Bol86}
L.~Boltzmann, Phil.\ Mag.\ {\bf 22}, 226 (1886).

\bibitem{Ada15}
E.P.~Adams, Proc.\ Am.\ Phil.\ Soc.\ {\bf 54}, 47 (1915).

\bibitem{Gal91}
For a~comprehensive review of early-stage researches, see:
S.~Galdamini and G.~Giuliani, Ann.\ Sci.\ {\bf 48}, 21 (1991).

\bibitem{Kir94}
G.~Kirczenow, J.\ Phys.: Condens.\ Matter {\bf 6}, L583 (1994);
S.~Souma and A.~Suzuki, Phys.\ Rev.\ B {\bf 58}, 4649 (1998).

\bibitem{Man96}
R.G.~Mani, Europhys.\ Lett.\ {\bf 36}, 203 (1996).

\bibitem{Ryc99}
S.F.W.R.\ Rycroft {\em et al.}, Phys.\ Rev.\ B {\bf 60}, 757(R) (1999).

\bibitem{Yan10}
J.~Yan and M.S.~Fuhrer, Nano Lett.\ {\bf 10}, 4521 (2010).

\bibitem{Fau10}
C.~Faugeras {\em et al.}, ACS Nano {\bf 4}, 1889 (2010).

\bibitem{Zha12}
Y.~Zhao {\em et al.}, 
Phys.\ Rev.\ Lett.\ {\bf 108}, 106804 (2012).

\bibitem{Kha13}
Z.~Khatibi, H.~Rostami, and R.~Asgari, Phys.\ Rev.\ B {\bf 88}, 195426 (2013).

\bibitem{Two06}
J.~Tworzyd{\l}o {\em et al.}, Phys.\ Rev.\ Lett.\ \textbf{96}, 246802 (2006).

\bibitem{Pra07}
E.~Prada {\em et al.}, 
Phys.\ Rev.\ B \textbf{75}, 113407 (2007).

\bibitem{Ryc09}
A.~Rycerz, P.~Recher, and M.~Wimmer, Phys.\ Rev.\ B \textbf{80}, 125417 (2009).

\bibitem{Ryc12}
A.~Rycerz, Acta Phys.\ Polon.\ A {\bf 121}, 1242 (2012).

\bibitem{voltfoo}
We assume the inner (or the outer) lead is characterized by the electrochemical potential $\mu_0-eV_{\rm eff}/2$ (or $\mu_0+eV_{\rm eff}/2$); the actual source-drain voltage may differ from $V_{\rm eff}$ due to charge-screening effects. 

\bibitem{Naz09}
Yu.V.~Nazarov and Ya.M.~Blanter, {\it Quantum Transport: Introduction to Nanoscience,} Cambridge University Press (Cambridge, 2009). 

\bibitem{Abram}
M.~Abramowitz and I.A.~Stegun, eds., 
\textit{Handbook of Mathematical Functions} (Dover Publications, Inc., New York, 1965), Chapter 13.

\bibitem{bsignfoo}
Without loss of generality, we choose $B>0$. For $B<0$ one gets $T_j(B)=T_{-j}(-B)$.

\bibitem{Kol12}
E.B.~Kolomeisky, H.~Zaidi, and  J.P.~Straley, 
Phys.\ Rev.\ B {\bf 85}, 073404 (2012). 

\bibitem{Sin11}
A.~Singha {\em et al.}, Science {\bf 332}, 1176 (2011).

\bibitem{Liu14}
Z.K.~Liu {\em et al.}, \url{dx.doi.org/10.1126/science.1245085}.

\bibitem{Bor14}
S.~Borisenko {\em at al.}, \url{arXiv:1309.7978} (unpublished).

\bibitem{mu0foo}
We notice that this supposition does not affect the universality of the results. For any $eV_{\rm eff}>2|\mu_0|$, in the high-field limit, the leading contributions to averages in Eqs.\ (\ref{ffinvo}) and (\ref{rfinvo}) originate from a~small vicinity of the Dirac point. The same reasoning applies to higher LLs. 

\bibitem{Rut14}
G.~Rut and A.~Rycerz, Phys.\ Rev.\ B \textbf{89}, 045421 (2014). 

\bibitem{fooinfty}
We have fixed the voltage at $eV_{\rm eff}R_{\rm i}/(\hbar{}v_F)=0.5$ for the data presented. No statistically signifficant effects were detected for other $V_{\rm eff}$-s  in the $1/m_\Phi\rightarrow{}0$ limit.

\end{thebibliography}
\end{document}